\documentclass{article}

\usepackage{arxiv}
\usepackage{paralist}
\usepackage{booktabs}
\usepackage{multirow}
\usepackage{listings}
\usepackage{subcaption}
\usepackage{xcolor}
\usepackage{pifont}
\usepackage{amsmath}
\usepackage{amsthm}
\theoremstyle{definition}

\usepackage{ragged2e}
\usepackage{amssymb}
\usepackage{footnote}
\usepackage{enumitem}
\usepackage{url}
\usepackage{comment}
\usepackage{tabularx}
\usepackage{array}
\usepackage{booktabs}
\usepackage{adjustbox}
\usepackage{tikz}
\usepackage{amsmath}
\usepackage[affil-it]{authblk}

\makesavenoteenv{tabular}
\makesavenoteenv{table}

\def\WithComments{}
\ifdefined \WithComments
\newcommand{\zhc}[1]{\textcolor{blue}{\textbf{#1}}}
\else
\newcommand{\zhc}[1]{}
\fi

\newcommand{\tab}{\hspace*{1em}}
\newcommand{\code}[1]{{\fontfamily{cmtt}\fontseries{m}\fontshape{n}\selectfont\small{#1}}}

\newcommand{\Romannum}[1]{\uppercase\expandafter{\romannumeral #1\relax}}

\newcommand{\system}{\code{Conch}}
\newcommand{\normalsystem}{{Conch}}

\begin{document}

\title{Revisiting Challenges for Selective Data Protection of Real Applications}

\author[1]{Lin Ma}
\author[1]{Jinyan Xu}
\author[1]{Jiadong Sun}
\author[1]{Yajin Zhou\thanks{Corresponding author (yajin\_zhou@zju.edu.cn).}\hspace*{0.4em}}
\author[1]{Xun Xie}
\author[1]{Wenbo Shen}
\author[1]{Rui Chang}
\author[1]{Kui Ren}
\affil[1]{Zhejiang University}
 
 \maketitle
 	
\begin{abstract}
\label{sec:abstract}
Selective data protection is a promising technique to defend against the data leakage attack. In this paper, we revisit technical challenges that were neglected when applying this protection to real applications. These challenges include the secure input channel, granularity conflict, and sensitivity
conflict. We summarize the causes of them and propose corresponding solutions. Then we design and implement a prototype system for selective data protection and evaluate the overhead using the RISC-V Spike simulator. The evaluation demonstrates the efficiency (less than 3\% runtime overhead with optimizations) and the security guarantees provided by our system.

\end{abstract}

\section{Introduction}\label{sec:introduction}

Selectively protecting sensitive data is a promising technique
to defend against the data leakage attack. 
Some recent systems~\cite{carr2017datashield,brahmakshatriya2019confllvm,palit2019mitigating,yun2019ginseng}
implement this type of protection to improve the performance by only protecting
the sensitive data instead of all memory objects. 
To achieve this, they require the developer to annotate variables that may contain
sensitive data and then use the static analysis tool to find all potential
candidates, e.g., variables copied from the sensitive data, that need to be protected. DataShield~\cite{carr2017datashield} and ConfLLVM~\cite{brahmakshatriya2019confllvm}
prepare dedicated memory regions and additional bound checking for potentially sensitive variables. 
Ginseng~\cite{yun2019ginseng} protects sensitive data by always putting it into registers except when task switching or interrupts occur.

Although these approaches are different in details, they all use the static data flow analysis to find out possible memory locations that may hold the sensitive data. Then they instrument all load and store instructions that could operate on these locations to achieve the protection.
However, this methodology suffers from the following shortcomings. 
First, the static analysis has a precision issue
in the points-to analysis~\cite{hind2001pointer} when locating sensitive data. 
Second, the protection granularity is coarse-grained. When protecting a sensitive candidate
inside a data structure, they need to mark the entire structure as sensitive,
introducing unnecessary performance overhead.

To resolve these shortcomings and achieve practical protection, we introduce \system{}, a solution to
selectively safeguard the confidentiality of data against diverse data leakage attacks.
\textit{The goal is to ensure that the sensitive data is never exposed to
untrusted user-space memory as plaintext for its entire lifetime.}
To this end, 
\system{} leverages the dynamic information flow tracking~\cite{chow2004understanding}
technique supported by underlying hardware (tagged architecture) to offer
precise selective data protection, thus solving the imprecise static point-to analysis problem.
Also, the protection is in the machine-word-level granularity, which is more fine-grained
than the structure-level granularity.
Besides, \system{} utilizes the hardware supported in-memory cryptographic transformation with per-thread encryption keys to offer robust confidentiality protection.
With the protection of our system, the attacker cannot spoil the confidentiality of
the sensitive data even under a strong threat model since the leaked data is encrypted.

However, when applying this solution to real applications, we need to confront three challenges \textit{that
were usually neglected by previous systems}. First, the sensitive data can
be from external sources such as a file. The data could be stored by the OS kernel into
a user-provided buffer when using the system call (e.g., the \code{read} system call)
to read the content from the file.
Thus, a mechanism is needed to mark the sensitive data inside the buffer. 
Second, the granularity conflict between the hardware tag and the memory manipulation
instructions (\code{load} and \code{store}) needs to be solved. Otherwise,
the tag for sensitive data could be cleared, compromising the security
guarantees of sensitive data protection.
Third, the sensitivity conflict will cause semantic compatibility of the program. We need to find a
solution to maintain the compatibility, while at the same time, cannot introduce the security loophole
that could be abused by the attacker to bypass our protection.

We implement a system prototype on the RISC-V simulator with new instructions, tagged registers, and
cache. We also implement the tag prorogation inside the CPU pipeline.
We use the mibench~\cite{guthaus2001mibench} and the real-world applications to analyze the performance overhead when protecting sensitive data. 
After that, we perform a security analysis to evaluate the security guarantees provided by our system.
The result shows that our system  can safeguard the confidentiality of sensitive data with a
performance overhead that is less than 3\%.

\smallskip \noindent
\textbf{Contributions}\tab
In summary, our work makes the following main contributions.
\begin{compactitem}
    \item We propose and implement a system to efficiently safeguard the confidentiality
    of the sensitive data against diverse data leakage attacks with a strong threat model.
    
    \item We summarize three technical challenges that were usually neglected by previous systems
    when applying selective data protection to real-applications, and propose corresponding strategy to confront them.
    
    \item We evaluate the overhead and security guarantees provided by our system.
    The result shows that \system{} can successfully defend against sensitive data leakage attacks with an appropriate performance overhead. 
  
\end{compactitem}

\section{Background}
\label{sec:background}

\subsection{Sensitive Data Leakage Attack}
The attack uses vulnerabilities in programs to exfiltrate sensitive data from the memory. In the following, we present typical scenarios that lead to sensitive data leakage.

\smallskip 
\noindent
\textbf{Format String Vulnerability}\tab Attackers could exploit the format string vulnerability~\cite{shankar2001detecting} 
to gain arbitrary memory read and write primitive. Even though many tools are presented to detect this vulnerability, it is still exists in real-world applications~\cite{cve20196840,cve202015203} nowadays.

\noindent
\textbf{Buffer Over-read}\tab It is a type of vulnerability where a program, while reading data from a buffer, overruns the buffer's boundary and reads (or tries to read) adjacent memory~\cite{strackx2009breaking}. 
For instance, the HeartBleed vulnerability allows the attacker to over-read around 64KB data from the memory buffer adjacent to the allocated buffer for the heartbeat packet. 
Besides, existing exploits use vulnerabilities like the type confusion, off-by-one, and the integer overflow to corrupt the metadata, especially the length of a dynamic object to achieve the buffer over-read primitive.

\subsection{Tagged Architecture}
The tagged architecture, proposed in 1973~\cite{feustel1973advantages}, equips the machine memory with additional tags.
Since its debut, it has been leveraged for security hardening, e.g., dynamic information flow tracking (DIFT)~\cite{suh2004secure,newsome2005dynamic,kannan2009decoupling,kang2011dta++,enck2014taintdroid,song2016hdfi}, memory safety~\cite{woodruff2014cheri,gallagher2019morpheus},  instrumentation and debugging~\cite{greathouse2012case}, 
capability protection~\cite{kwon2013low,woodruff2014cheri}, 
and others~\cite{shrobe2009trust,dhawan2014pump,song2015towards}. 
For example, Minos~\cite{crandall2004minos} uses one-bit tags to indicate the integrity of code pointers. CHERI~\cite{woodruff2014cheri} also uses one-bit tags to decide if an address stores a valid capability. 
The Morpheus~\cite{gallagher2019morpheus} uses two-bit domain tags to distinguish between code, code pointers, data pointers and other data. 
Our work also leverages the tagged architecture for runtime sensitive data tracking.

\section{Motivating Example and Threat Model}
\label{sec:example}

\smallskip 
\noindent
\textbf{The HeartBleed Vulnerability}\tab
The HeartBleed (CVE-2014-0160) is a serious vulnerability in the OpenSSL library. It is a buffer over-read bug in the implementation of the heartbeat extension, which allows the attacker to leak sensitive data from the remote server.
The code snippet in Fig.~\ref{fig:heartbleed} shows the
vulnerable function (\code{tls1\_process\_heartbeat()}).

\begin{figure}[t]
	\adjustbox{scale = 1}{
		\centering
		\begin{lstlisting}[basicstyle=\bfseries\fontsize{7}{7}\ttfamily, keywordstyle=\bfseries\color{blue}\ttfamily, stringstyle=\color{red}\ttfamily, commentstyle=\color{brown}\ttfamily, language=c, breaklines=true, tabsize=2,  numbers=left, firstnumber = 1, gobble=4]
			// p is the pointer to incoming packet 
			n2s(p,payload);    
			pl=p;
			buffer=OPENSSL_malloc(1+2+payload+padding);
			bp=buffer;
			*bp++=TLS1_HB_RESPONSE;
			s2n(payload,bp);
			memcpy(bp,pl,payload); // over-read at pl!
			r = ssl3_write_bytes(s,TLS1_RT_HEARTBEAT,buffer,3 + payload+padding);

	\end{lstlisting}
	}
	\caption{The vulnerable \code{tls1\_process\_heartbeat()} function in OpenSSL version 1.0.1f.}
	\label{fig:heartbleed}
\end{figure}

Specifically, the vulnerable function allocates a buffer (line 4) with the payload
size (\code{payload}) extracted from the heartbeat request packet (line 2), which is controlled by the attacker.
It finally constructs the heartbeat response packet using this buffer (line 10 to 12). 
The attacker can craft a malicious heartbeat request packet with a large value at the \code{payload} field while only appending a small size of payload. 
Despite the size of the buffer for storing the actual requesting packet (what \code{pl} points to) is much less than the value of the \code{payload}, the memory copy is still executed, resulting in the buffer over-read bug. As a result, the sensitive data in memory could be leaked to attacker. To defend this, our system ensures that sensitive data is \textit{always encrypted} in the memory, with the encryption key maintained by the kernel. The vulnerability may still exist. However, it cannot be exploited to leak the private data in the plaintext.

\smallskip 
\noindent
\textbf{Threat Model and Assumptions}\tab
Our system assumes a strong threat model that the attacker can read arbitrary use-space memory.
This can be achieved through the nature of the vulnerability (the HeartBleed).
However, the attacker cannot inject code into execution due to the availability of DEP.
Also the attacker cannot hijack the control flow of the program.
These assumptions align with previous works~\cite{carr2017datashield,brahmakshatriya2019confllvm,palit2019mitigating,yun2019ginseng}.
We do not consider the exploit of kernel vulnerabilities to leak user-space memory. Also,
the side-channel attack and the cold-boot attack are out of the scope.

\section{Overall System Design}
\label{sec:design}


Our work aims to \textit{selectively protect sensitive data by ensuring that it is never exposed to untrusted user-space memory as plaintext ever since its initialization}.
To this end, our system requires the developer to annotate the source of sensitive data. 
Then the compiler wraps the source with new CPU instructions to enable tag initialization.
When the program executes on the CPU, it utilizes the tagged architecture to dynamically track the propagation of the sensitive data inside the registers and caches. 
Before any tagged data being written into the memory, an encryption engine will transparently encrypt the data. As a result, the sensitive data will never be leaked as plaintext in the user-space memory. In the following, we will illustrate the main steps of our system.

\smallskip 
\noindent
\textbf{Annotating Sensitive Data}\tab
\system{} selectively tracks and encrypts the developer-specified sensitive data.
By focusing only on this subset, the performance overhead can be limited.
Moreover, this human-in-the-loop strategy can be 
flexible and accurate compared to an automatic one~\cite{enck2014taintdroid}.

Most of the previous systems~\cite{carr2017datashield,brahmakshatriya2019confllvm,yun2019ginseng}
take the type-based annotation and allow the developer to mark the definition of memory objects in the source code to regard them as \textit{sensitive}. 
However, the type-based annotation cannot support fine-grained protection because the granularity of the sensitive data remains in the data structure level. 
For \system{}, developers just need to add annotations at the sensitive data starting point, aka, \textit{sensitive sources}. 
The compiler will automatically generate the machine instructions to mark the data as sensitive (using the memory tag.)
Though the idea of annotating is rather simple, it is not trivial to implement because the sensitive data can come from multiple sources, such as files in the disks, inputs from keyboards, and random bytes from the pseudo-random number generator. The challenge will be discussed in Section~\ref{sec:challenges}.

\smallskip 
\noindent
\textbf{Propagating Tags}\tab
In our system, a one-bit tag is applied to a machine word. As a result, it only imposes
1.56\% memory space overhead on modern 64-bit architecture. 
The memory accesses will be split into the data access and the tag access.
To decrease the incurred extra DRAM traffic overhead, the tag cache optimization
is adopted, which is proven to be highly useful~\cite{song2015towards,joannou2017efficient}.

The tagged architecture is similar to systems for information flow tracking~\cite{bradbury2014tagged,gallagher2019morpheus}.
The architecture associates each data with its tag in both the execution pipeline and
the memory hierarchy.
To offer robust protection, it introduces tag (or taint) propagation rules 
that enable a lifetime tracking for the sensitive data as well as its transformed variants. 
The rules strictly ensure all operations that involve sensitive data should propagate
the sensitivity to the result. 
When the CPU executes a memory unrelated instruction,
the tag of the source register may propagate to the destination register,
according to the propagation rules. 
For memory related instructions such as \code{store} and \code{load}, the tags
should be stored into, or loaded from the memory together with the data. 
To facilitate this process, our system augments the registers and caches to hold the tags.

\smallskip 
\noindent
\textbf{Encrypting Data}\tab
Our system encrypts the sensitive data (tagged data) before being written into the memory.
With the memory tag, the encryption engine is transparent to the program
as it does not require any instrumentation to distinguish sensitive data from others.

Our system utilizes a strong block cipher named QARMA~\cite{avanzi2017qarma}.
Compared with the commonly used AES, the major advantage of QARMA  is that
it enables an additional input named \code{tweak} to parameterize the permutations.
For each memory word, our system chooses its address as the tweak used in the encryption. 
Hence the same sensitive data in different addresses will have different encrypted outcomes.

Our system has fine-grained management of the encryption keys by the operating system kernel.
Once the system is booting, each CPU processor randomly generates a master key. 
After that, when a new thread is created, a per-thread key will be generated
using this master key and then be associated with the context of this thread. 
Our system offers two additional micro-architectural registers that cannot
be accessed from user-space to store the master key and the currently in-use thread key.
The master key register preserves the master key ever since its initialization. 
The thread key register, however, keeps changing as the thread is actively scheduling. 
During the context switch, the old thread's key will be saved on the kernel stack, and
the current thread's key is loaded.

\section{Practical Challenges}
\label{sec:challenges} 
Though the architecture is straightforward and similar to the previous one~\cite{gallagher2019morpheus},
applying such a defense on real programs needs to confront several challenges (These challenges were usually ignored by previous systems~\cite{roessler2018protecting,gallagher2019morpheus,palit2019mitigating}. In this section, we will present the details of these challenges and the coping strategies.

\subsection{Sensitive Input Channel}
\label{subsec:sensitive_input}

The sensitive data is not allowed to be exposed in the user-space memory without encryption. However, the sensitive data may already reside inside the memory before the developer has the chance to annotate it, e.g., writing by the system call in the OS kernel. In other words, our system should pay attestation to every possible channel through which the sensitive data enters into the user-space memory (\textit{Sensitive Input Channel} in this paper).

Previous systems either omit this problem~\cite{palit2019mitigating} or require the
developer to rewrite the code and ensure the buffer that receives the sensitive data
is protected via isolation~\cite{carr2017datashield}, which requires additional engineering effort.
Our system proposes a design of sensitive input channel that can associate the
tag to the sensitive data and initialize the defense before the data being written into userspace memory.
The construction of this channel is not trivial due to the following two reasons.
First, the sensitive data may come from various sources, such as files in the disks,
keyboards, network sockets. Second, the sensitive input channel should not change the
way (or the APIs) that are used to maintain the program's compatibility.

Specifically, \system{} leverages the OS kernel to build the sensitive input channel. 
This is because the kernel is in charge of processing the data from the sensitive source
and placing it into the supplied user-space memory buffer. 
However, the kernel has no idea whether the incoming buffer contains sensitive data or not.
A direct solution is to use dedicated devices for sensitive inputs like Ginseng~\cite{yun2019ginseng}, which
needs extra UART devices. However, it does not satisfy the issues of multiple types of input sources.

Our system solves the issue by allowing the developers to inform the kernel
that the passing buffer will contain sensitive inputs.
To accomplish this, it patches existing system calls in the kernel. 
For instance, a file is opened with the specified \code{O\_SENSITIVE}
flag so the kernel can return a file descriptor associated with the additional attribute. 
When the program reads data from this special file descriptor, the kernel can switch to the
sensitive input channel and initialize the protection before the data is placed into the user-space buffer.

In the implementation,
\system{} patches the \code{copy\_to\_user()} function, which is an interface that transmit the data from kernel-space to user-space memory.  
This function will check if the file descriptor is opened with the \code{O\_SENSITIVE} flag to perform the actual data transmission. 

Besides, our system changes the \code{getrandom} system call, which is widely used in cryptographic algorithms, to give random value that is protected with tags.

\subsection{Sensitivity Conflict}
\label{subsec:sensitivity}
Another challenge is the sensitivity conflict that 
will 
cause the behaviors of program to not match the expectation. 
One example is when protecting the session key of the SSL/TLS handshaking process
in the OpenSSL library, the content that is encrypted using this session key
(in the \code{tls1\_enc()} function) will be tagged as sensitive, due to the tag prorogation.
Thus, it will be encrypted again by \system{} when writing into
the memory before being sent to the client. Hence, the received data cannot be decrypted since the client does
not have the encrypted key maintained by \system{}.

To solve this challenge, we provide an instruction that can remove the memory tag of a buffer. 
The developer can insert this instruction
to the program when the data is already protected (encrypted) and needs to be shared
with another entity (SSL client for instance.) 
However, we need to ensure that this instruction cannot be abused by the attacker
to remove the tag of arbitrary memory. 
Fortunately, the control flow of the
program cannot be hijacked (See the threat model in Section \ref{sec:example}), thus the attacker
cannot redirect the control flow to remove the tag of a buffer controlled by the attacker.

\subsection{Granularity Conflict}
\label{subsec:granularity}
\begin{figure}[t]
	\centering
	\includegraphics[width=0.7\linewidth]{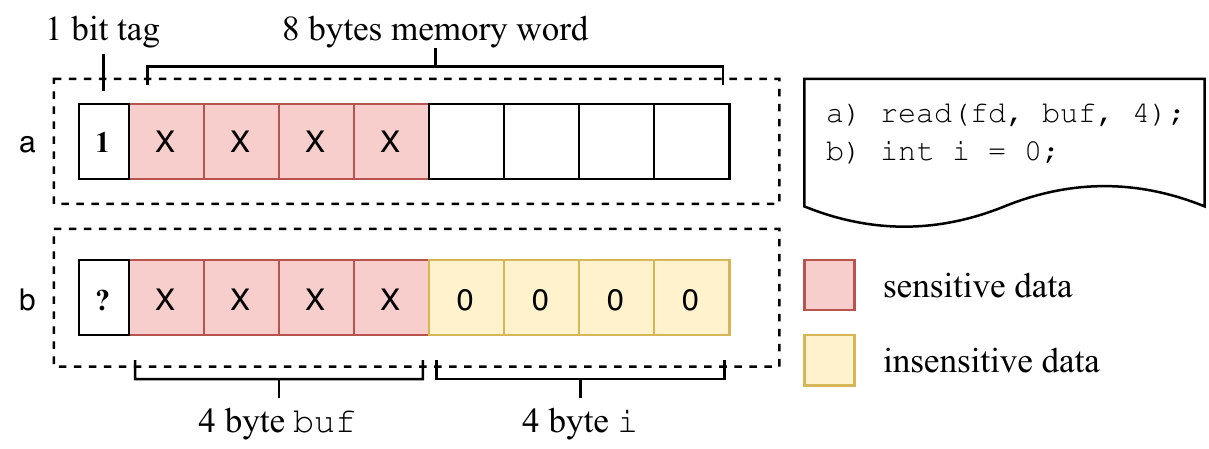}

	\caption{The granularity conflict: The dotted frames represent the state of memory word state after the a and b lines of code. \textbf{?} marks the uncertain tag value.}

	\label{fig:gconflict}
\end{figure}

\system{} has one-bit tag for a machine word. This may introduce the granularity conflict between a memory tag and the byte granularity of the memory load and store instructions.
We use the example in Figure.~\ref{fig:gconflict} to demonstrate this issue. 
As the figure shown, there are two variables (a 4-byte array \code{buf} and 4-byte integer \code{i}) placed into one machine word by the compiler. 
The code \textbf{a} reads four bytes sensitive data from a dedicated file descriptor into the \code{buf} so the memory tag will be set to 1 because the associated machine word contains sensitive data. 
However, after executing the code \textbf{b} which puts a constant value into the variable \code{i}, the ambiguity about the value of the tag of this memory word is introduced since the machine word now contains both the sensitive data and normal data.

The direct solution for this conflict is to extend the tagged architecture to support per-byte memory tags and ensure one tag will be necessarily associated with exactly one variable,
However the per-byte tags leads to a leap of memory overhead from the 1.56\% to 12.5\% to store the tags. Though there is architecture~\cite{dhawan2014pump} claims to support unbounded bits of tag, most tagged architectures~\cite{suh2004secure,kwon2013low,woodruff2014cheri,song2016hdfi} utilize only one-bit tags to reduce the memory overhead.

After surveying the recent works that utilize the memory tag for security, we sadly find that this conflict is rarely discussed. Some researches~\cite{song2016hdfi,gallagher2019morpheus} only focused on word aligned objects, like pointers. Nick's work~\cite{roessler2018protecting} adopts the tag to protect data in the stack. But they only discuss word-aligned variables and ignore the fact the unaligned variables could exist. 

\system{} firstly discusses the impact of granularity conflict. To ensure the sensitive data is always protected, we decide to retain the tag and according to our evaluation (Section \ref{subsec:granularity}), the extra overhead from over-tagging is acceptable.

\section{System Implementation}
We have implemented a prototype system 
based on the Spike simulator by extending the pipeline and registers to construct the tagged architecture. 
Because the vanilla Spike lacks the simulation of cache, we write our own cache module plugin to simulate the instruction and data cache. 
We implement the extended instructions, based on the RISC-V custom instructions support, into the Spike and modify the v9.2.0 GCC compiler for support.
Besides, we implement the QARMA algorithm~\cite{avanzi2017qarma}, i.e., $\text{QARMA}_5\text{-}64\text{-}\sigma_1$, which encrypts 64-bits block data with a 128-bits key in 12 rounds. To accurately model the memory system and assess the overhead of tag accessing, we utilize the state of art memory simulator DRAMSim3~\cite{li2020dramsim3}. 
Last, we add the key management and the designed sensitive input channels in the v5.4.7 Linux kernel. 

\section{Evaluation}
\label{sec:evaluation}

\subsection{Methodology}
\label{eval_method}

We adopt a methodology used in TIMBER-V~\cite{Weiser2019} to estimate the system
runtime performance overhead by mapping executed instructions into actual cycles
using different pipelined CPU models. 
To this end, we rewrite the histograms module in the Spike simulator to allow the
precise recording of all executed instructions and trace all memory accesses. 
Specifically, we first define a baseline model without the tagged memory extension and then compare with another two models, one of which is equipped with tag cache optimization and another one is not. 
Different from TIMBER-V, which assigns cycles for each instruction empirically, we build a more accurate pipelined model and configure the CPU and cache referring to the SiFive CPU manual~\cite{sifive2018}. 
We additionalhy utilize the memory simulator DRAMSim3 and add the tag cache into Spike to evaluate the memory access overhead.

\smallskip
\noindent \textbf{Baseline CPU Model}\tab
As a baseline, we configure the simulator referring to the Freedom FU540-C000 Soc.
Based on that, we model the instructions cycle according to the SiFive manual. For the memory \code{load} and \code{store} instructions, they cost two or three cycles depending on operand values when the cache hits. 
When the cache misses, we model the latency according to the DRAMSim3 simulator.

\smallskip
\noindent  \textbf{\normalsystem{} CPU Model}\tab
There are two models, namely \normalsystem{} Model A and \normalsystem{} Model B.
The Model B is optimized with tag cache while A is not.
For the encryption engine of these two models, we assume a four-cycles latency for $\text{QARMA}_5\text{-}64\text{-}\sigma_1$ to encrypt or decrypt a block, similar to the previous work~\cite{gallagher2019morpheus,liljestrand2019pac}.
For the optimized Model B, we configure the 4KB eight-way associative tag cache.

\subsection{Benchmarks}

\begin{figure}[t]
	\centering
	\includegraphics[width=0.7\linewidth]{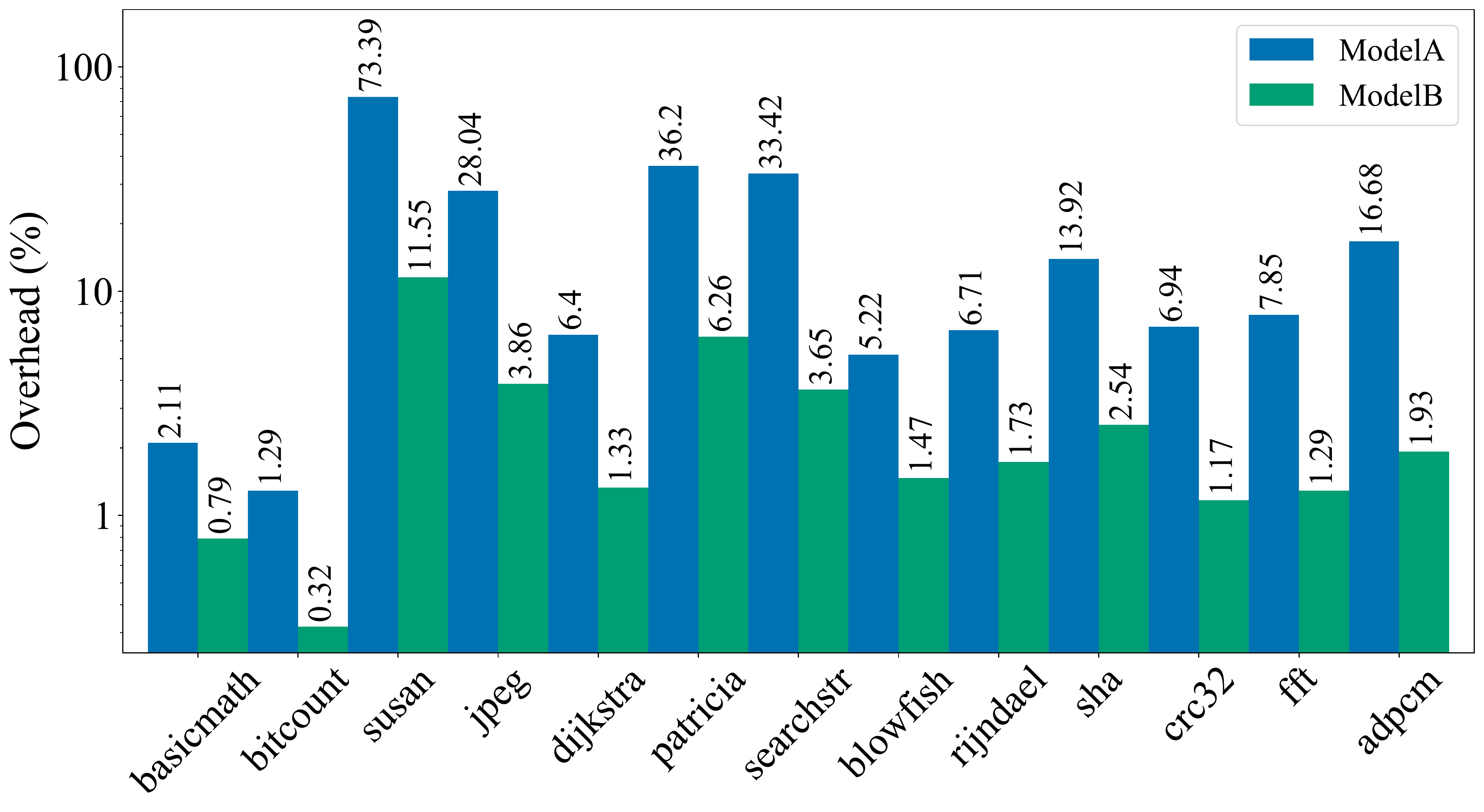}

	\caption{The runtime overhead of Conch Model A and Model B (with tag cache support) for mibench~\cite{guthaus2001mibench}}

	\label{fig:mibench}
\end{figure}

To estimate the cost caused by the tagged architecture, we use the
mibench~\cite{guthaus2001mibench} to measure the overhead.
Since we do not annotate any sensitive data, the overhead is all caused by the tag architecture. 
The result is shown in Fig.~\ref{fig:mibench}. This result shows an average runtime overhead of $18.3\%$ for the Model A with the maximum value of $73.39\%$ for the program {susan}, an image processing program that performs large amounts of memory accesses. 
Other programs with frequent memory access have a comparatively high overhead compared with the CPU-intensive programs. 
However, with the tag cache optimization, the result ($2.914\%$ on average) is promising. It shows that the tag cache can efficiently decrease the runtime overhead of tagged architecture.

\subsection{Real Programs}
\label{subsec:app}
\begin{figure}[t]
	\centering
	\includegraphics[width=0.7\linewidth]{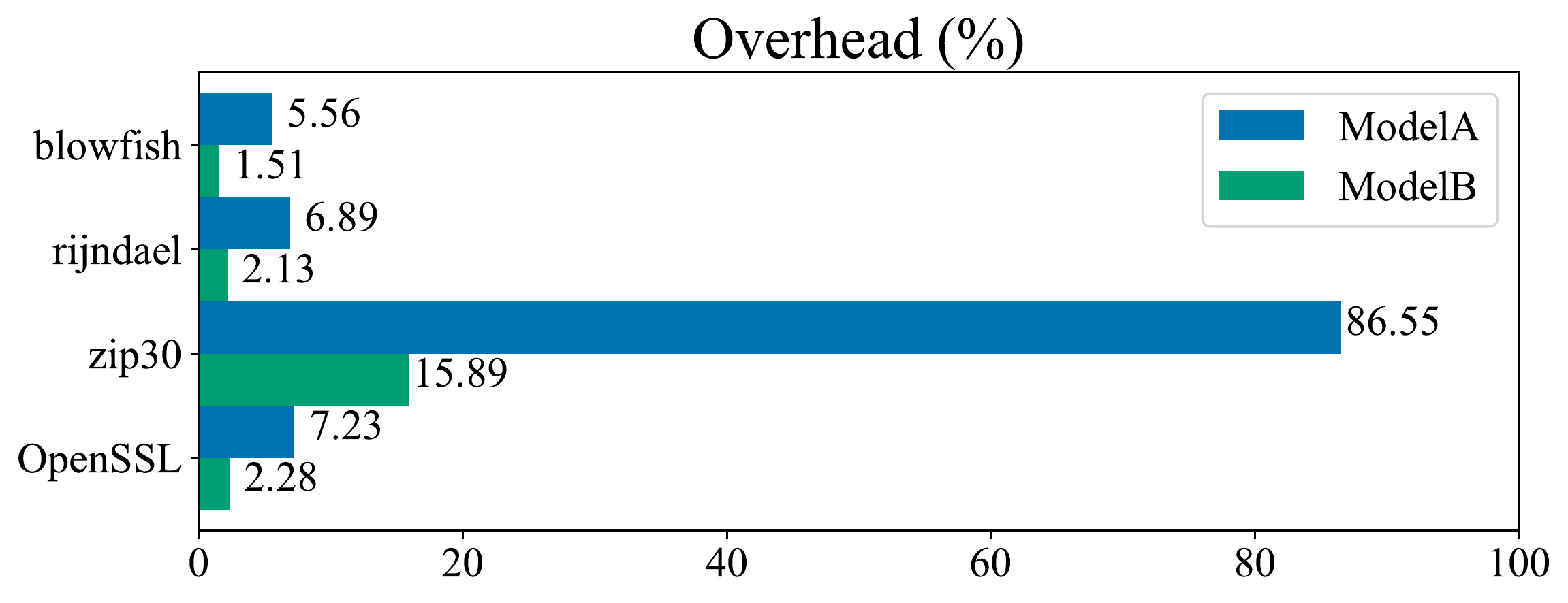}

	\caption{The runtime overhead of Conch Model A and Model B (with tag cache support) in real-world application when protecting sensitive data.}

	\label{fig:casestudy}
\end{figure}

In this section, we use \system{} to protect the sensitive data in real-world programs
and analyze the performance overhead.
We choose the cryptographic applications including blowfish and rijndael from the mibench as targets.
Besides, we select the {zip30} application and the OpenSSL library to enrich the evaluation.
The overall result is shown in Fig.~\ref{fig:casestudy}.

\smallskip
\noindent \textbf{Cryptographic Algorithms}\tab
To protect the sensitive data (encryption keys) of the cryptographic algorithms
{blowfish} and {rijndael}, 
we annotate the source code to ensure that the password, which is involved in the generation of the encryption key, is protected once it enters the memory.
We rewrite these two programs to read the password from the sensitive file with
the \code{O\_SENSITIVE} flag. 
As the Fig.~\ref{fig:casestudy} show, our system has low performance overhead. For the Model A, \system{} imposes 5.56\% and 6.89\% overhead for {blowfish} and {rijndael}, respectively. For Model B, the overhead lower to 1.51\% and 2.13\%. 

\smallskip
\noindent \textbf{Cryptographic Applications}\tab
We use the {zip30} and the OpenSSL as target applications.
The first one is a widely used compression utility, which implements a protection to encrypt
the file content using a user-provided password. 
To protect the sensitive password, we rewrite the program to load the password from a specialized TTY input channel. 
The performance overhead is 86.55\% for Model A and 15.89\% for the optimized Model B, respectively.
This relatively high overhead comes from the high frequency of memory accesses as it trivially reads uncompressed input data chunks and writes encrypted output to memory.

For the OpenSSL, we use \system{} to protect its SSL/TLS communications. 
Besides the sensitive master key and the session key, we also protect the private key
used in handshaking. 
To do this, we rewrite the \code{SSL\_CTX\_use\_PrivateKey\_file()} function to allow it to read the sensitive private key from the specialized channel our system supplies. 
For the master key and the session key, they are both obtained from the pseudo-random generator function \code{tls1\_PRF()}.
Hence, we annotate this sensitive source to initialize the protection. 
The result (7.23\% for Model A and 2.27\% for Model B) shown in Fig.~\ref{fig:casestudy} indicates the high efficacy even when protecting the sensitive data in a relatively complex application.

\subsection{Granularity Conflict Study}
\label{subsec:granularity}

As discussed in Section~\ref{subsec:granularity}, the choice of retaining the tag will cause
over-tagging issue. We show the experiment result of the over-tagging ratio in Table~\ref{tab:gresult}.

\begin{table}[t]
    \centering
    \caption{The result of granularity conflict of 4 applications. In the table, we show the percentage of
    sensitive data that has been over-tagged and extra runtime overhead for each application.}
    \label{tab:gresult}
    \resizebox{.7\textwidth}{!}{
    \begin{tabular}{|c|c|c|c|c|} 
    \hline
     & blowfish & rijndael & zip30 & OpenSSL \\
    \hline
    Over-tagging & 0.033\%/0.14\% & 0\%/0\% & 0.13\%/1.04\% & 5.12\%/2.16\% \\
    \hline
    \end{tabular}
    }
\end{table}

The result shows that except rijndael, three other applications suffer from the granularity conflict. 
In particular, there is 0.033\% sensitive data over-tagging for blowfish. 
That said, 0.033\% of the protected data is supposed to be insensitive.
The over-tagging result is 0.13\% for zip30, and 5.12\% for OpenSSL.
We additionally estimate the cost of protecting the wrongly tagged data. 
The result is it imposes 0.14\%, 1.04\%, and 2.16\% extra runtime overhead for
blowfish, zip30, and OpenSSL, respectively.

\subsection{Security Analysis}
\system{} aims to safeguard the confidentiality of sensitive data.
The threat model is strong as the attacker is granted with arbitrary read and write primitive.
Because \system{} utilizes the tagged architecture to dynamically track the sensitive data and encrypt it before it is written into memory, the sensitive data cannot be leaked.

In the following, we will analyze the possible loophole due to the granularity and sensitivity conflict.

\smallskip
\noindent \textbf{Granularity Conflict}\tab
Our system retains the tag in the granularity conflict. Thus it will not remove tags
for sensitive data. As a result, the security guarantees still hold, although unnecessary
data may be protected.

\smallskip
\noindent \textbf{Sensitivity Conflict}\tab
We allow the developer to rectify the tag propagation with the ability to remove the memory tag of a buffer to solve the sensitivity conflict issue (Section~\ref{subsec:sensitivity}).
In our threat model, the attacker cannot hijack the control flow so the tag removing instruction cannot be abused to bypass our protection.

\section{Related Work}
\label{sec:relatedwork}
\smallskip
\noindent \textbf{Selective Data Protection}\tab Instead of protecting all memory objects, selective data protection only focuses on partial targets, such as pointers, control-flow related variables, and the developer specified sensitive data. 
Some recent approaches achieve this by requiring the programmer to annotate sensitive memory objects in the source code.
DataShield\cite{carr2017datashield} and ConfLLVM\cite{brahmakshatriya2019confllvm} prepare dedicated memory region and additional bound checking for the annotated sensitive data.
Tapti et al\cite{palit2019mitigating}, similar to \system{}, selectively encrypts the sensitive content before it is written into memory. 
However, the cryptographic transformations they designed are done by the instrumentation code instead of the hardware engine hence has poor performance.
Ginseng\cite{yun2019ginseng} innovatively protects annotated sensitive data by allocating them to registers at compile time and only puts it into memory when task switching or interrupts occur.
To defend against a compromised kernel, Ginseng leverages the Trusted Execution Environment (TEE) to manage encryption and decryption. 
These approaches bear disadvantages like high runtime overhead, coarse granularity, and the difficulty of deploying.

\smallskip
\noindent \textbf{Tag-based Memory Protection}\tab
Several systems utilize the tagged architecture for protection end. 
For example, 
lowRISC\cite{kwon2013low} uses tags to specify whether a memory address is readable or writable. 
HDFI\cite{song2016hdfi} uses tags to achieve data-flow integrity. 
CHERI\cite{woodruff2014cheri} uses tags to indicate if a memory address stores a valid fat pointer. 
Among the previous work, Morpheus\cite{gallagher2019morpheus} is closest to \system{}. 
It uses two-bit domain tags to distinguish code, code pointers, data pointers, and normal data and further adopts the domain encryption defense to randomizes the representation of code, code pointers, and data pointers before they are written into memory. 
The difference is that \system{} aims to protect the developer specified sensitive data, which is the superset of Morpheus's target. 
Additionally, \system{} promises protection ever since the data is initialized and only consumes one-bit tags, comparatively more lightweight and easy to implement.

\section{Conclusion}
In this paper, we revisit the technical challenges that were usually neglected by
previous systems when applying selective data
protection to real-applications, and propose corresponding solutions.
Then we design and implement a prototype system for selective data protection
and evaluate the overhead using the Spike simulator. The evaluation demonstrates
the efficiency (less than 3\% overhead with optimizations) and the security guarantees
provided by our system.

\bibliographystyle{plain}
\bibliography{main}

\end{document}